# Controller synthesis with very simplified linear constraints in PN model

Dideban A. * Zareiee M. * Alla H. **

* *Semnan University, IRAN (Tel:* (+98)231-3354123*; e-mail: adideban@ Semnan.ac.ir).*
* *Semnan University, IRAN (Tel:* (+98)231-3354123*; e-mail:mzareiee@ Semnan.ac.ir).*
*** GIPSA lab, 38402 St Martin d'Heres Cedex FRANCE , Hassane.alla@inpg.fr).*

**Abstract:** This paper addresses the problem of forbidden states for safe Petri net modeling discrete event systems. We present an efficient method to construct a controller. A set of linear constraints allow forbidding the reachability of specific states. The number of these so-called forbidden states and consequently the number of constraints are large and lead to a large number of control places. A systematic method for constructing very simplified controller is offered. By using a method based on Petri nets partial invariants, maximal permissive controllers are determined.

## 1. INTRODUCTION

A unifying framework for the control of discrete event systems is provided by Ramadge and Wonham (1987, 1989), and so far their supervisory control theory is the most general and comprehensive theory presented for logical DES. In Ramadge & Wonham (1987, 1989), it is shown that we can model the systems by using of formal languages and finite automata. Methods exist for designing controllers based on automata system models, however these methods often involve exhaustive searches or simulations of the system behaviour, making them impractical for systems with large numbers of states and transitions causing event (Moody, Antsaklis 2000). For solving these problems, we can model systems with Petri nets (PNs).

Petri nets are a very appropriate and useful tool for the study of discrete event systems because of their modelling power and their mathematical properties. In the last decade, the research in the field of controller synthesis of discrete event systems has become one of the most active domains (Achour et al. 2004; Giua A & Xie X., 2005).

In Giua et al., (1992) it is shown that it is possible to represent the forbidden states with linear constraints. With the idea presented by Yamalidou et al. (1996) we can design a controller by adding a control place to the PN mode. Each control place corresponds to a constraint. However, when the number of forbidden states is large, then the number of constraints has the same size. In Giua (1992b) it is shown that we can simplify some constraints. But there is no systematic method for simplifying these constraints in general case. This simplification is done by using the PNs structural properties.

In Dideban and Alla (2005) a systematic method for reducing the size and the number of constraints was introduced. It was applicable for safe and conservative PNs.

In Dideban and Alla (2008), an efficient method for constructing a controller based on safe PN is presented. A set of linear constraints allow forbidding the reachability of specific states. The number of these so-called forbidden states, and consequently the number of constraints are large and lead to a large number of control places. A systematic method for reducing the size and the number of constraints for safe PN was introduced there.

With applying the idea in Dideban & Alla (2008), it is possible that sometimes the final number of constraints may remian large. In this paper our objective is to develop the idea in Dideban & Alla (2008) and to present a systematic method for safe PN which allows to do more simplification than in the previous approaches.

In this paper, we introduce the concept of Partial Place invariant and then, we present the simplification method. We apply this method to a concrete example to highlights its advantages. We use constraints which are equivalent to forbidden states. These constraints can be calculated in two different ways. They can be given directly as specifications or they can be deduced thanks to the Kumar approach (Kumar & Holloway, 1996).

By applying the idea in Dideban & Alla (2008) and the new approach, it is possible to make a powerful simplification. The advantage of the new idea is to reduce more the number of forbidden states.

This paper is oraganized as follows: in section 2, some important definitions are introduced. In section 3, the idea of passage from forbidden states to linear constraints will be introduced. The idea in Dideban & Alla (2008) and a practical example are presented in section 4. The basic original idea of simplification and the calculation of control places are presented in section 5. Finally, the conclusion is given in the last section.

## 2. PRELIMINARY PRESENTATION

In this paper, it is supposed that the reader is familiar with the PNs basis (David & Alla, 2005) and the theory of supervisory control (Ramadge & Wonham, 1987, 1989). We briefly introduce the idea that was presented in (Dideban & Alla, 2008). For more detail, the reader can refer to this reference. .

A PN is represented by a quadruplet $R=\{P, T, W, M_0\}$ where $P$ is the set of places, $T$ is the set of transition, $W$ the incidence matrix and $M_0$ is the initial marking. Here we show the number of tokens in a place $P_i$ with $m_i$. The PN is assumed to be safe; then the marking of each place is Boolean.

$\mathcal{M}_R$ denotes the set of PN reachable markings. In $\mathcal{M}_R$, two subsets can be distinguished; the set of authorized states $\mathcal{M}_A$ and the set of forbidden states $\mathcal{M}_F$. The set of forbidden states correspond to two groups:

1) The set of reachable states ($\mathcal{M}_F'$) which either don't respect the specifications or are deadlock states.

2) The set of states for which the occurrence of uncontrollable events leads to states in $\mathcal{M}_F'$.

The set of authorized states are the reachable states without the set of forbidden states:

$$\mathcal{M}_A=\mathcal{M}_R\backslash\mathcal{M}_F$$

Among the forbidden states, an important subset is constituted by the border forbidden states denoted as $\mathcal{M}_B$ (Kumar & Holloway, 1996).

*Definition 1.* let $\mathcal{M}_B$ be the set of border forbidden states:

$$\mathcal{M}_B=\{M_i\in\mathcal{M}_F \mid \exists\sigma\in\textstyle\sum_c \text{ and } \exists M_j\in\mathcal{M}_A, M_j \xrightarrow{\sigma} M_i\}$$

Where $\sum_c$ is the set of controllable transitions. ❑

In this paper we use the concept of support instead of marking as presented in definition 2.

*Definition 2***.** the function Support($X$) of a vector $X\in\{0,1\}^N$ is: Support($X$) = The set of marked places in $X$. ❑

*Definition 3.* let $\mathcal{R}$ be a PN , $\mathcal{P}$ the set of its places , $\mathcal{M}(M_0)$ the set of reachable markings from $M_0$ and k a constant. a place invariant is obtained if there is a sub-set of places $\mathcal{P}'=\{P_1, P_2,\ldots, P_r\}$ included in P and a weighing vector $(q_1, q_2,\ldots, q_r)$ for which all the weights $q_i$ are positive integers such that:

$$q_1m_1+q_2m_2+\ldots+q_rm_r=k,\ \forall\ m_i\in\mathcal{M}(M_0)$$

Then we say $\mathcal{P}'$ is a place invariant, (David & Alla 2005). ❑

But if we remove some places from the set $\mathcal{P}'$, what happens? It follows the definition of partial invariant:

*Definition 4***.** let $\mathcal{P}'=\{P_1,P_2,\ldots,P_r\}$ be a place invariant in PN model $\mathcal{R}$, $\mathcal{P}_{i1}=\{P_1,P_2,\ldots,P_L\}$ for which $\{1,2,..L\}\subset\{1,2,..r\}$, is a partial place invariant and it satisfy the following equation:

$q_1m_1+q_2m_2+\ldots+q_Lm_L\le k$ , $\forall m\in\mathcal{M}(M_0)$ ❑

## 3. FORBIDDEN STATES TO LINEAR CONSTRAINTS

Let $M_i(M_i^T=[m_1,m_2,\ldots,m_N])$ Be a forbidden state in set $\mathcal{M}_B$ and Support($M_i$)=$\{P_{i1}P_{i2}P_{i3}\ldots P_{in}\}$ the set of marked places of $M_i$. From a forbidden state, a linear constraint can be constructed (Giua et al., 1992).

The linear constraint deduced from the forbidden state $M_i$ is given below. The state $M_i$ does not verify this relation. Therefore, by applying this relation, $M_i$ will be forbidden.

$$\sum_{k=1}^{n} m_{ik}\le n-1$$

Where n=Card [support ($M_i$)], is the number of marked places of $M_i$, and $m_{ik}$ is the marking of place $P_{ik}$ of state $M_i$.

Let $M$ ($M^T=[m_1,m_2,\ldots,m_N]$) be a general marking and $M_i$ be a forbidden state. The constraint (forbidden state $M_i$) is denoted by $c_i$ and can be rewritten as follows

$$M_i^T.M\le \text{Card}[\text{support}(M_i)]-1$$

For example if:

$$M_i^T=[0,1,1,0,0,0,1]\Rightarrow \text{Card}[\text{support}(M_i)]=3$$
$$\Rightarrow m_2+m_3+m_7\le 2$$

Verifying the relation in above is equivalent to forbidding state $M_i$ when the PN model is conservative. However, in a safe PN not necessarily conservative, this equivalence is not always true.

In follow we recall the concept of over state that was introduced in Dideban & Alla (2008).

*Definition 5.* let $M_2=P_{21}P_{22}\ldots P_{2m}$ be an accessible state, $M_1=P_{11}P_{12}\ldots P_{1n}$ will be an over-state of $M_2$ if:

$M_1\le M_2$ .

That means all of the places in $M_1$ are in $M_2$. ❑

For example $M_1=P_1P_3$ is an over-state of $M_2=P_1P_3P_6P_9$.

The name over-state is used because the constraint corresponding to an over-state holds constraint. For example, the constraint $m_4+m_6\le1$ that is correspond to the over-state $M_1=P_4P_6$ holds both following constraints:

$$m_1+m_4+m_6+m_9\le3 \quad , \quad m_2+m_4+m_6+m_9\le3$$

These two constraints forbid states $M_6=P_1P_4P_6P_9$ and $M_7=P_2P_4P_6P_9$. $P_4P_6$ is an over-state of both states $P_1P_4P_6P_9$ and $P_2P_4P_6P_9$ which could be verified by $M_1\le M_6$ and $M_1\le M_7$. Thus by using only the constraint $m_4+m_6\le1$, both states $M_6$ and $M_7$ will be forbidden.

*Remark 1***.** With each over-state $b_i$, we associate a constraint $c_i$ : $b_i=(P_{i1}P_{i2}P_{i3}\ldots P_{in})\rightarrow c_i=(P_{i1}P_{i2}P_{i3}\ldots P_{in},n\text{-}1)$.

That means: $m_{i1}+m_{i1}+\ldots+m_{in}\le n\text{-}1$. ❑

The concept of over-state could be used for the reduction the number of constraints deduced from states or over-states (Dideban & Alla 2008). In the next section we introduce these ideas and the problem to solve.

## 4. RECUCTION THE NUMBER OF CONSTRAINTS

Now we explain the idea in Dideban & Alla (2008). Suppose that in a system we have the set of authorized states $\mathcal{M}_A$ and the set of border forbidden states $\mathcal{M}_B$. In the first step of reduction, we must construct the sets $\mathcal{B}_1$ (All the over-states of border forbidden states) and $\mathcal{A}_1$ (All the over-states of authorized states). Then we remove from $\mathcal{B}_1$, the states or over-states that exist in $\mathcal{A}_1$ and call this set $\mathcal{B}_2$. In the next step we remove the states which their over-states are in $\mathcal{B}_2$ from it and we call it $\mathcal{B}_3$.

Sometimes it is possible, to simplify the reduced set $\mathcal{B}_3$, by application of the idea that is presented in Dideban (2007). But sometimes it is not the simplest set. The rules to choose the final over-states are similar to the rules of the Quine-McCluskey method to simplify the logical expressions (Morris Mano 2001).

### *4.1. Introduction of a practical example*

Consider a system with three machines and two robots. The closed loop PN model of this system is presented in fig. 1. The start command of each machine is accomplished by controllable transitions $c_1$, $c_2$, $c_3$ and the stop command is accomplished by uncontrollable transitions $f_1$, $f_2$, $f_3$. Two Robots are used to free three machines. All of machines can be turned on or off independently. The end of process for each machine when at least one of the robots are not free, is forbidden.
From constructing the reachability graph, we can calculate the set of border forbidden states.

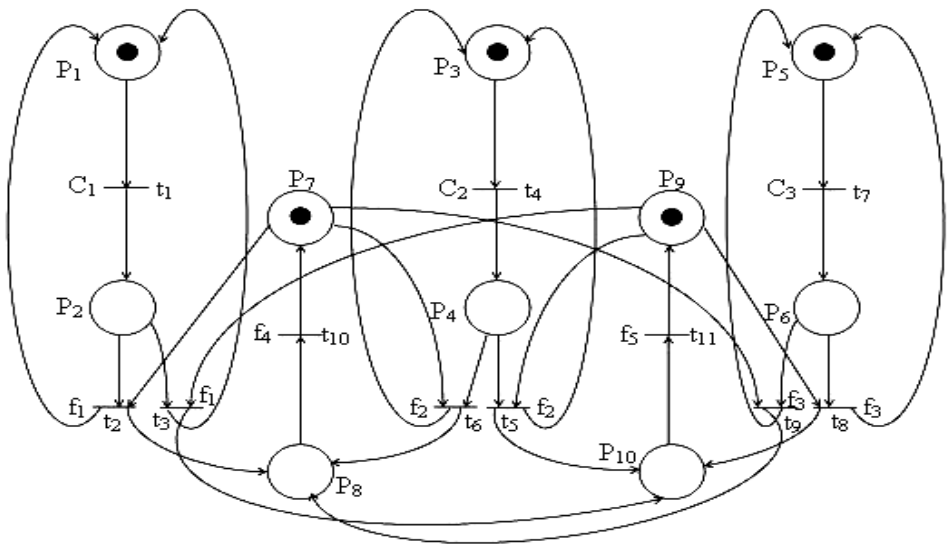

Fig. 1. closed loop PN model

Fig.2, gives part of reachability graph of the PN model presented in fig 1. The construction of the reachability graph is stopped when a forbidden state is reached. The state after these forbidden states is presented by $\otimes$ .

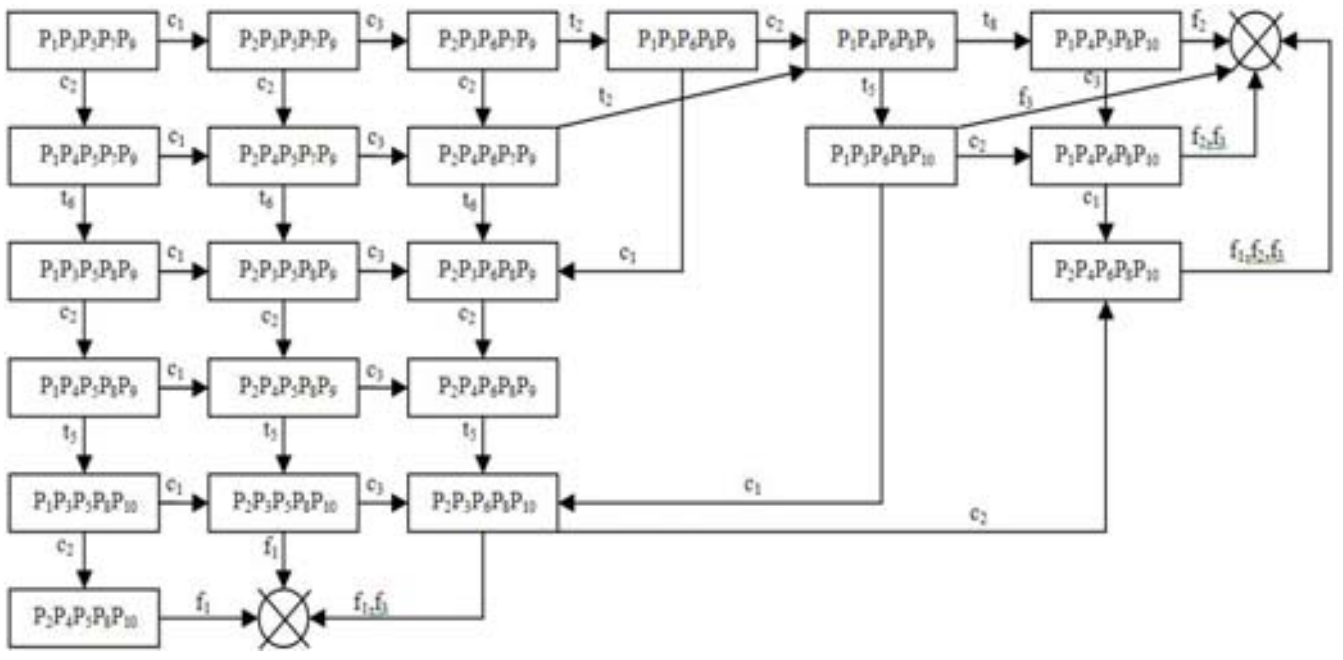

Fig. 2. A part of reachability graph of the PN model

As it is obvious in fig. 2, the existence of uncontrollable events leads to the forbidden states. For example when the system is in state ($P_1P_4P_5P_8P_{10}$), it is possible to fire the uncontrollable event $f_1$, while it is not authorized by the specifications. This state is called as a forbidden state. And the set of border forbidden states can be determined from these forbidden states.
In this example the set of border forbidden states is:

$$\mathcal{M}_B = \{p_2p_4p_6p_7p_9, p_2p_4p_6p_8p_9, p_2p_4p_6p_7p_{10}, p_2p_4p_6p_8p_{10}, p_2p_4p_5p_8p_9, p_2p_4p_5p_7p_{10}, p_2p_3p_6p_8p_9, p_2p_3p_6p_7p_{10}, p_1p_4p_6p_8p_9, p_1p_4p_6p_7p_{10}, p_2p_3p_5p_8p_{10}, p_1p_4p_5p_8p_{10}, p_1p_3p_6p_8p_{10}\}$$

And The set of authorized states is:

$$\mathcal{M}_A = \{p_1p_3p_5p_7p_9, p_1p_3p_5p_8p_9, p_1p_3p_5p_7p_{10}, p_1p_3p_5p_8p_{10}, p_2p_3p_5p_7p_9, p_2p_3p_5p_8p_9, p_2p_3p_5p_7p_{10}, p_1p_4p_5p_7p_9, p_1p_4p_5p_8p_9, p_1p_4p_5p_7p_{10}, p_1p_3p_6p_7p_9, p_1p_3p_6p_8p_9, p_1p_3p_6p_7p_{10}, p_2p_4p_5p_7p_9, p_2p_3p_6p_7p_9, p_1p_4p_6p_7p_9\}$$

Now, the sets of over-states of $\mathcal{M}_A$ and $\mathcal{M}_B$ are constructed. The set of over-states of $\mathcal{M}_B$ is shown with $\mathcal{B}_1$ and $\mathcal{M}_A$ with $\mathcal{A}_1$.
Hence the set of over-states of $\mathcal{M}_B$ is:

$$\mathcal{B}_1=\{p_1,p_2,p_3,p_4,p_5,p_6,p_7,p_8,p_9,p_{10},p_2p_4,p_2p_6,p_2p_7,p_2p_8,p_2p_9,p_2p_{10},p_2p_3,p_2p_5,p_1p_8,p_1p_9,$$
$$p_1p_7,p_1p_{10},p_1p_4,p_1p_6,p_1p_5,p_1p_3,p_3p_6,p_3p_8,p_3p_9,p_3p_7,p_3p_{10},p_3p_5,p_4p_6,p_4p_7,p_4p_8,p_4p_9,p_4p_{10},$$
$$p_4p_5,p_5p_8,p_5p_9,p_5p_{10},p_5p_7,p_6p_7,p_6p_8,p_6p_9,p_6p_{10},p_7p_9,p_7p_{10},p_8p_9,p_8p_{10},p_1p_3p_6,p_1p_3p_8,p_1p_3p_{10},$$
$$p_1p_8p_{10},p_1p_6p_7,p_1p_6p_{10},p_1p_7p_{10},p_1p_4p_5,p_1p_5p_8,p_1p_5p_{10},p_1p_8p_{10},p_1p_4p_6,p_1p_4p_7,p_1p_4p_{10},p_1p_6p_7,$$
$$p_1p_6p_{10},p_1p_4p_7,p_1p_4p_{10},p_2p_4p_8,p_2p_6p_8,p_2p_8p_9,p_2p_4p_6,p_2p_4p_7,p_2p_4p_9,p_2p_6p_7,p_2p_6p_9,p_2p_7p_9,$$
$$p_2p_4p_{10},p_2p_6p_{10},p_2p_7p_{10},p_4p_6p_{10},p_2p_8p_{10},p_2p_4p_5,p_2p_5p_8,p_2p_5p_9,p_2p_5p_7,p_2p_5p_{10},p_2p_3p_6,p_2p_3p_8,$$
$$p_2p_3p_9,p_2p_3p_7,p_2p_3p_{10},p_2p_3p_5,p_3p_6p_8,p_3p_6p_9,p_3p_8p_9,p_3p_6p_7,p_3p_6p_{10},p_3p_7p_{10},p_3p_8p_{10},p_3p_5p_8,$$
$$p_3p_5p_{10},p_4p_6p_8,p_4p_8p_9,p_4p_6p_7,p_4p_6p_9,p_4p_7p_9,p_4p_7p_{10},p_4p_8p_{10},p_4p_5p_8,p_4p_5p_9,p_4p_5p_7,p_4p_5p_{10},$$
$$p_5p_8p_9,p_5p_8p_{10},p_5p_7p_{10},p_6p_8p_9,p_6p_7p_9,p_6p_7p_{10},p_6p_8p_{10},p_1p_3p_6p_8,p_1p_3p_6p_{10},p_1p_3p_8p_{10},p_1p_6p_8p_{10},$$
$$p_1p_4p_5p_8,p_1p_4p_5p_{10},p_1p_4p_8p_{10},p_1p_5p_8p_{10},p_1p_6p_7p_{10},p_1p_4p_6p_7,p_1p_4p_6p_{10},p_1p_4p_7p_{10},p_1p_6p_7p_{10},p_2p_4$$
$$p_6p_7,p_2p_4p_6p_9,p_2p_4p_7p_9,p_2p_6p_7p_9,p_2p_4p_6p_{10},p_2p_4p_5p_9,p_2p_5p_8p_9,p_2p_4p_7p_{10},p_2p_6p_7p_{10},p_2p_4p_8p_{10},$$
$$p_2p_6p_8p_{10},p_2p_4p_5p_8,p_2p_3p_6p_8,p_2p_3p_6p_9,p_2p_3p_8p_9,p_2p_4p_5p_7,p_2p_4p_5p_{10},p_2p_5p_7p_{10},p_2p_3p_6p_7,p_2p_3p$$
$$_6p_{10},p_2p_3p_7p_{10},p_2p_3p_5p_8,p_2p_3p_5p_{10},p_2p_3p_8p_{10},p_2p_5p_8p_{10},p_2p_4p_6p_8,p_2p_4p_8p_9,p_2p_6p_8p_9,p_3p_6p_8p_9,$$
$$p_3p_6p_7p_{10},p_3p_5p_8p_{10},p_3p_6p_8p_{10},p_4p_6p_8p_9,p_4p_6p_7p_{10},p_4p_6p_8p_{10},p_4p_5p_8p_9,p_4p_5p_7p_{10},p_4p_5p_8p_{10}\}$$

After this step we remove from $\mathcal{B}_1$, the states that are common in $\mathcal{B}_1$ and $\mathcal{A}_1$, and the residual set is named $\mathcal{B}_2$.
*Remark 2***.** we don't write the set of over-states of authorized states $\mathcal{A}_1$ since with having $\mathcal{M}_A$ we can remove the common states from $\mathcal{B}_1$. ❑

$$\mathcal{B}_2 = \{P_2P_4P_6, P_2P_4P_8, P_2P_6P_8, P_4P_6P_8, P_2P_4P_{10}, P_2P_6P_{10}, P_4P_6P_{10}, P_2P_8P_{10}, P_6P_8P_{10}, P_4P_8P_{10}\}$$

By the method presented in Dideban & Alla (2008), the controller can be synthesized by final selection of simplified over-states.

## 4.2. Controller synthesis

Final selection is similar to the McCluskey method for simplifying logical terms. In this method we construct a table (table1) where the first row represents the set of border forbidden states $\mathcal{M}_B$, and the first column is the set of simplified over-states $\mathcal{B}_2$. The relation between border forbidden states and the simplified over-states can be expressed by the following definition:
*Definition 6.* let $\mathcal{B}_3=\{b_1,b_2,\dots,b_m\}$ be the set of simplified over-states and $\mathcal{M}_B=\{M_1,M_2,\dots,M_N\}$ be the set of border forbidden states. The relation $R: M_B \times B_3 \to \{0,1\}$ is as:

$$R(M_i,b_j)=\begin{cases}1 & b_j \le M_i (b_j \; is\; over-state\; of\; M_i)\\ 0 & if\; not.\end{cases}$$

The covering of a marking is an integer number:

$$Cv(M_i)=\sum R(M_i,b_j)$$

$Cv(M_i) \geq 1$ means that forbidden state $M_i$ is covered by at least one over-state. ❑

**Table 1: Relation between over-states and border forbidden states**

| | $P_2P_4P_6P_7P_9$ | $P_2P_4P_6P_8P_9$ | $P_2P_4P_6P_7P_{10}$ | $P_2P_4P_6P_8P_{10}$ | $P_2P_4P_5P_8P_9$ | $P_2P_4P_5P_7P_{10}$ | $P_2P_3P_6P_8P_9$ | $P_2P_3P_6P_7P_{10}$ | $P_1P_4P_6P_8P_9$ | $P_1P_4P_6P_7P_{10}$ | $P_2P_3P_5P_8P_{10}$ | $P_1P_4P_5P_8P_{10}$ | $P_1P_3P_6P_8P_{10}$ |
|---|---|---|---|---|---|---|---|---|---|---|---|---|---|
| $P_2P_4P_6$ | 1 | 1 | 1 | 1 | 0 | 0 | 0 | 0 | 0 | 0 | 0 | 0 | 0 |
| $P_2P_4P_8$ | 0 | 1 | 0 | 1 | 1 | 0 | 0 | 0 | 0 | 0 | 0 | 0 | 0 |
| $P_2P_6P_8$ | 0 | 1 | 0 | 1 | 0 | 0 | 1 | 0 | 0 | 0 | 0 | 0 | 0 |
| $P_4P_6P_8$ | 0 | 1 | 0 | 1 | 0 | 0 | 0 | 0 | 1 | 0 | 0 | 0 | 0 |
| $P_2P_4P_{10}$ | 0 | 0 | 1 | 1 | 0 | 1 | 0 | 0 | 0 | 0 | 0 | 0 | 0 |
| $P_2P_6P_{10}$ | 0 | 0 | 1 | 1 | 0 | 0 | 0 | 1 | 0 | 0 | 0 | 0 | 0 |
| $P_4P_6P_{10}$ | 0 | 0 | 1 | 1 | 0 | 0 | 1 | 0 | 0 | 1 | 0 | 0 | 0 |
| $P_2P_8P_{10}$ | 0 | 0 | 0 | 1 | 0 | 0 | 0 | 0 | 0 | 0 | 1 | 0 | 0 |
| $P_4P_8P_{10}$ | 0 | 0 | 0 | 1 | 0 | 0 | 0 | 0 | 0 | 0 | 0 | 1 | 0 |
| $P_6P_8P_{10}$ | 0 | 0 | 0 | 1 | 0 | 0 | 0 | 0 | 0 | 0 | 0 | 0 | 1 |
| $C_v(M_j)$ | 1 | 4 | 4 | 10 | 1 | 1 | 2 | 1 | 1 | 1 | 1 | 1 | 1 |
| | | | | | | | | | | | | | |

The final constraints must cover all the border forbidden states. This means that with applying the constraints equivalent to the final sets, all the border forbidden states are forbidden. These constraints are as follow:

$$\mathcal{B}_3 = \{P_2P_4P_6, P_2P_4P_8, P_2P_4P_{10}, P_4P_6P_{10}, P_2P_6P_{10}, P_4P_6P_8, P_2P_8P_{10}, P_4P_8P_{10}, P_6P_8P_{10}\}$$

Before using the method in Dideban & Alla (2008), the number of constraints and equivalently the number of border forbidden states for our example was 13, but with using this method, the number of constraints is 9. It is clear that the number of simplified constraints is large, and for designing a controller by the method presented by Yamalidou et al. (1996), the number of control places is also large. However, is it possible to simplify these constraints more? The answer to this question is given by a method based on the concept of partial place invariant that was defined in section 2.

## 5. INTRODUCTION OF THE NEW IDEA FOR SIMPLIFICATION

We consider the partial invariant for safe and conservative PNs. From the invariant relations we can construct partial invariants. For example:

$$m_1+m_2+\ldots+m_n= k \quad \Rightarrow \quad m_1+m_2+\ldots+m_{n-1}\leq k$$

In safe PN in general case, it is not possible to construct this equation directly. In the following, we present some properties based on partial invariants that in many cases allow us to simplify the set of constraints representing the forbidden states.

*Property 1.*; suppose that $m_{i1}$, $m_{i2}$ are the number of tokens in the places $P_{i1}$, $P_{i2}$ respectively. If $P_{i1}P_{i2}$ is not in the set of over-states of authorized states we have:

$m_{i1}+m_{i2}\leq 1$

❑

**Proof:** Suppose that the constraint is not true. Then we have:

$m_{i1} + m_{i2} > 1$

So, for safe PNs we have:

$$m_{i1}+m_{i2}=2 \rightarrow m_{i1}= m_{i2}=1$$

This means $P_{i1}P_{i2}$ is in the set of over-states of authorized states that it is not true. Then: $m_{i1}+m_{i2}\leq 1$

❑

*Property 2 (Extension of Property 1)***.** suppose that the constraint $m_{i1}+\ldots+m_{in}\leq 1$ is true. If all of the over- states $\{P_{i1}P_{i(n+1)},\ldots,P_{in}P_{i(n+1)}\}$ are not in the set of over-states of authorized states, we have: $m_{i1}+\ldots+m_{in}+m_{i(n+1)}\leq 1$ ❑

**Proof:** The proof of this property is similar to the proof of property1. Suppose that this relation is not true, so we can write:

$$m_{i1}+\ldots+m_{in}+m_{i(n+1)} >1$$

hence we have:

$$\left.\begin{array}{l} m_{i1}+\ldots+m_{in}+m_{i(n+1)} = 2 \\ m_{i(n+1)}\leq 1 \text{ (for safe PN)} \\ m_{i1} +\ldots+ m_{in} \leq 1 \end{array}\right\} \Longrightarrow \left\{\begin{array}{l} m_{i(n+1)} = 1 \\ m_{i1} +\ldots+ m_{in} = 1 \end{array}\right.$$

$$m_{i1} +\ldots+ m_{in} = 1 \quad \Rightarrow \quad \exists\, m_{ik} = 1 (\, k \in [1,n])$$

Then $P_{ik}P_{i(n+1)}$ is an over-state of authorized states that is not true, then: $m_{i1} + \ldots+ m_{in} + m_{i(n+1)} \leq 1$ ❑

*Remark 3***.** A constraint $m_1 + m_2 +\ldots+ m_k \leq k$ can be presented by $(P_1P_2\ldots P_k , k)$. ❑

*Property3***.** Let $\mathcal{M} = \{(P_1P_k \ldots P_j, k),\ldots,(P_r P_k \ldots P_j, k)\}$ be the subset of constraints verified the authorized states. If the authorized states verify the partial invariant $m_1 + m_2 +\ldots + m_r \leq 1$ the $r$ constraints are equivalent to one constraint as bellow:

$$( m_1 + m_2 +\ldots+ m_r) + m_k +\ldots+ m_j \leq k$$ ❑

**Proof:** *Necessary condition*:

$$\forall\, q \in \{1,\ldots, r\} \quad m_q+ m_i +\ldots+ m_j \leq k \;\Rightarrow\; m_i +\ldots+ m_j \leq k$$

And also from the constraint, we have:

$$m_1 + m_2 +\ldots + m_r \leq 1$$

By adding two constraints, we have:

$$( m_1 + m_2 +\ldots+ m_r) + m_k +\ldots+ m_j \leq k +1$$

We show that the value $k$+1 is not any time accessible. Suppose that $k$+1 is accessible. We have:

$$m_1+m_2+\ldots+m_r=1 \;,\; m_i+\ldots+m_j= k$$

From the $r$ constraints and equation $m_i+\ldots+m_j=k$ , we have:

$$m_1=m_2=\ldots=m_r=0$$

Then: $(m_1+m_2+\ldots+m_r)=0$

That is not true.

*Sufficient condition:*

$$(m_1 +\ldots+ m_r) + m_i +\ldots+ m_j \leq k$$

$$\forall\, q \in \{1,\ldots, r\} \;\; m_q\geq 0 \;\;\Rightarrow$$

$$\forall\, q \in \{1,\ldots,r\} \qquad m_q + m_i +\ldots+ m_j \leq k$$

❑

Generally, by application of these properties, it is possible to simplify the constraints. However, for our example, it is not possible to do more simplification using these properties. Suppose that the sets $A'_1$ and $B'_1$ be the sets of authorized and forbidden states, respectively:

$$A'_1 = \{P_1P_3P_5P_7, P_2P_3P_5P_7, P_1P_4P_5P_8, P_2P_4P_6P_8\}$$

$$B'_1 = \{P_1P_4P_5, P_2P_4P_5, P_2P_5P_7, P_4P_5P_7\}$$

By application of property 1 and 3 for states $\{P_1P_4P_5, P_2P_4P_5\}$ we have: $m_1 + m_2 + m_4 + m_5 \leq 2$. For states $\{P_2P_5P_7, P_4P_5P_7\}$, the constraint $m_2 + m_4 \leq 1$ is not verified, but is it necessary? We show that it is not necessary. When the borne of constraints is more than 1, for having $m_2 + m_4 + m_5 + m_7 \leq 2$, it is not necessary $m_2 + m_4 \leq 1$, but also it is necessary that $m_2 + m_4 + m_5 \leq 2$ and $m_2 + m_4 + m_7 \leq 2$ be true.

The following propertie generalizes this idea.

*Property 4* suppose that we have the set of constraints:

$C = \{(P_{i1}P_1P_2\ldots P_n, n), (P_{i2}P_1P_2\ldots P_n, n)\}$. Consider $n$ over-states as follow:

$P_{i1}P_{i2}P_1P_2\ldots P_{j-1}P_{j+1}\ldots P_n$ for $2 \leq j \leq n-1$
$P_{i1}P_{i2}P_2\ldots P_n$ , $P_{i1}P_{i2}P_1P_2\ldots P_{n-1}$
If none of these states are not in the set of over-states of authorized states, then we can reduce these constraints to one constraint as follow: $P_{i1}P_{i2}P_1P_2\ldots P_n$ that means:

$m_{i1}+m_{i2}+m_1+m_2+\ldots+m_n \leq n$

❑

**Proof**: the proof of this property is clear. Suppose that the inequality $m_{i1}+m_{i2}+m_1+m_2+\ldots+m_n \leq n$ is not true then $m_{i1}+m_{i2}+m_1+m_2+\ldots+m_n = n+1$,, that means $(n+1)$ places is marked in the set of $\mathcal{P} = (P_{i1}, P_{i2}, P_1, P_2, \ldots, P_n)$ then at least one of the condition in the property is violated. So our supposition is not true. ❑

*Property 5*. Suppose that we have the constraints $C_1 = (P_{i1}P_{i2}\ldots P_{im}P_1P_2\ldots P_n, n)$ and $C_2 = \{(P_{i(m+1)}P_1P_2\ldots P_n, n)$, that verify the authorized states. If all of the over-states content $n+1$ places from the sets $\mathcal{P} = (P_{i(m+1)}, P_{i1}P_{i2}\ldots P_{im}P_1P_2\ldots P_n,)$ are not exist in the sets of over-states of authorized states, the two constraints $C_1$ and $C_2$ can be replaced by one constraint as follow :

$P_{i1}P_{i2}\ldots P_{im}\, P_{i(m+1)}P_1P_2\ldots P_n, n$ ❑

**Proof:** the proof of this property is similar to the proof of property 4. Suppose that the inequality $m_{i1}+m_{i2}+\ldots+m_{im}+m_{i(m+1)}+m_1+m_2+\ldots+m_n \leq n$ is not true then $m_{i1}+m_{i2}+\ldots+m_{im}+m_{i(m+1)}+m_1+m_2+\ldots+m_n = n+1$, that means $(n+1)$ places is marked in the set of $\mathcal{P} = (P_{i1}, P_{i2}, \ldots, P_{im}, P_{i(m+1)}, P_1, P_2, \ldots, P_n)$ then at least one of the condition in the property is violated. So our supposition is not true.

*5.1. More simplification in our example by new idea*

From section 4.1 we have:

$$\mathcal{B}_3 = \{P_2P_4P_6, P_2P_4P_8, P_2P_4P_{10}, P_4P_6P_{10}, P_2P_6P_{10}, P_4P_6P_8, P_2P_8P_{10}, P_4P_8P_{10}, P_6P_8P_{10}\}$$

We can select the subset $C_1$ that contains two common places and one different place. $C_1 = \{(P_2P_4P_6, 2), (P_2P_4P_8, 2)$.

The over-state $P_6P_8$ is in the set of over-states of authorized states. So it is impossible to use the property 2, but the states $P_2P_6P_8$, $P_4P_6P_8$, are not in the set of over-states of authorized states, then it is possible to use the property 4.

$C_1 = \{(P_2P_4P_6, 2), (P_2P_4P_8, 2) \Rightarrow C'_1 = \{(P_2P_4P_6P_8, 2)$

For applying the property 5 on the sets of $C'_1$ and $C_2 = \{(P_2P_4P_{10}, 2)\}$, it is clear that the over-states $P_2P_8P_{10}$, $P_4P_8P_{10}$, $P_2P_6P_{10}$, $P_4P_6P_{10}$ are not in the set of over-states of authorized states, then, we arrive to the simplified constraint: $C'_2 = \{(P_2P_4P_6P_8P_{10}, 2)\}$

By the same way for the sets $C_3$ and $C_4$ we can arrive to the simplified constraints as bellow:

$C_3 = \{(P_2P_6P_{10}, 2), (P_4P_6P_{10}, 2), (P_6P_8P_{10}, 2)\} \Rightarrow$
$C'_3 = (P_2P_4P_6P_8P_{10}, 2)$

$C_4 = \{(P_2P_8P_{10}, 2), (P_4P_8P_{10}, 2), (P_4P_6P_8, 2)\} \Rightarrow$
$C'_4 = (P_2P_4P_6P_8P_{10}, 2)$
The simplified sets $C'_2$ and $C'_3$ and $C'_4$ are the same. Then the final set is: $C'_4 = (P_2P_4P_6P_8P_{10}, 2)$

So it is obvious that with using these properties the 9 constraints have reduced to one constraint that is good reason for utilization of these properties.

In this example, there is not the final selection, but generally by the method as the same was presented in last section, we can select the final result.

*5.2. Calculation of control places*

To calculate the control places corresponding for each linear constraint, we will use the method developed in Yamalidou et al. (2006). Now we briefly present this method that is based on concept of P-invariant. Suppose that we have the set of constraints as $L.m_p \leq b$ that $m_p$ is the marking vector of system, $L$ is a $n \times n$ matrix, $b$ is a $n_c \times 1$ vector, $n_c$ is the number of constraints and $n$ is the number of places. In this method for each constraint, we add a place to the PN model of the system. In Yamalidou et al. (2006), it is proved that this places guarantees that these constraints are respected. Suppose $W_p$ be the incidence matrix of the system. For each constraint we add a row to $W_p$ and we show this row as $W_c$, corresponds to the incidence matrix of the controller. and $W_c$ is calculated as follows:

$$W_c = -L.W_p$$

$W_c$ is added to $W_p$ as follow:

$$W = \begin{bmatrix} W_p \\ W_c \end{bmatrix}$$

If the initial marking of system is $m_{p0}$, the initial marking for the added places is calculated as follow:

$m_{s0}=b-L.m_{p0}$ $\qquad$ $m_0=\begin{bmatrix} m_{p0} \\ m_{s0} \end{bmatrix}$

Now we calculate the control places for our example. The simplified constraint was ($P_2P_4P_6P_8P_{10}$, 2), Then we have:
$L$=[0 1 0 1 0 1 0 1 0 1] $\Rightarrow W_c$=[-1 0 0 -1 0 0 -1 0 0 1 1], $M_{s0}$=2

The PN model of the final controller after adding the control place is indicated in fig 3. The control place and corresponding arcs are shown with gray color.

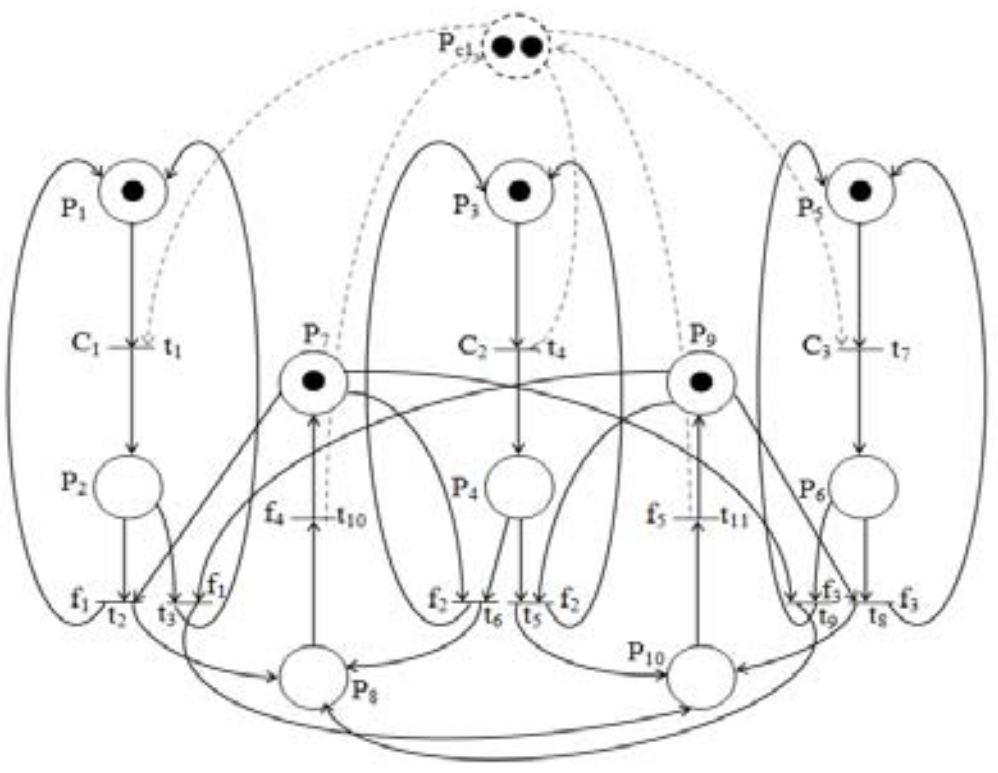

Fig. 3. Controlled PN model

As it is obvious from fig. 3, with adding $P_{c1}$, forbidden states can not be reached..

## 6. CONCLUSION

In this paper, we have presented a systematic method to reduce the number of linear constraints corresponding to the forbidden states for a safe PN. This is realized by using non-reachable states and by building the constraints using a systematic method. The important concept of over-state has been used; it corresponds to a set of markings which keep the same property (forbidden or authorized). From the forbidden states, the set of over-states is calculated. The utilization of non-reachable markings allows great simplification of the constraints.

In this paper, we have extended the idea presented in Dideban & Alla (2008). We have used the partial invariant for more simplifications. Finally, we have shown that sometimes it is possible to simplify without invariant or partial invariant. By this method we can arrive to very simplified constraints deduced from the forbidden states.

## REFERENCES

Achour Z., & N.Rezg, X. Xie . (2004). Supervisory controller of Petri nets under partial Observation. Proc. IFAC WODES04: 7th Workshop on DISCRETE EVENT SYSTEMS(Reims, france), September.

Dideban, A., & Alla, H. (2005). From forbidden state to linear constraints for the optimal supervisory control. *Control Engineering and appliedt Informaics* (*CEAI*), 7(3), 48-55.

Dideban A., 2007, Synthesis of discrete controllers by simplification of constraints and conditions, Ph.D. Thesis, UJF, France.

Dideban, A., & Alla, H. (2008). Reduction of Constraints for Controller Synthesis based on Safe Petri Nets. *Automatica,* 44(7): 1697-1706.

Giua, A., DiCesare, F.M., & Silva. (1992). Generalized Mutual Exclusion Constraints on Nets with Uncontrollable Transitions. In Proc. IEEE int. conf. on systems, man, and cybernetics (pp. 974–799).

Giua, A., & Xie, X. (2005). Control of safe ordinary Petri Nets using unfolding. Discrete Event Dynamic Systems: Theory and Applications, 15, 349–373.

Kumar, R., & Holloway, L.E. (1996). Supervisory control of deterministic Petri nets with regular specification languages. *IEEE Trans. Automatic Control*, 41(2):245-249.

Moody, J.O., & Antsaklis, P. (2000). Petri net supervisorfor DES with uncontrollable and unobservable transition", IEEE Trans. Automatic Control, 45(3): 462-476, mars.

Morris Mano, M. (2001). Digital design. Prentice Hall.

Ramadge, P.J., & Wonham,W.M. (1987). Supervisory control of a class of discrete event processes. *SIAM Joural of Control and Optimization,* 25 (1):206-230.

Ramadge, P. J., & Wonham, W. (1989). The control of discrete event systems.Dynamics of discrete event systems [Special issue]. Proceedings of theIEEE, 77(1), 81–98.

Rene, D., Alla, H. (2005). Discrete, Continuous, and Hybrid Petri Net. Springer.

Yamalidou, K., Moody, J., Lemmon, M., & Antsaklis, P. (1996). Feedback control of petri nets based on place invariants. Automatica, 32(1), 15–28.